\documentclass{iopjournal}
\usepackage[utf8]{inputenc}
\usepackage{amssymb}
\usepackage{amsmath}
\usepackage[mathscr]{euscript}
\usepackage{bbm}
\usepackage{bm}
\usepackage{mathtools}
\usepackage{graphicx}
\usepackage{natbib}
\usepackage{hyperref}
\hypersetup{
    colorlinks=true,
    linkcolor=blue,
    filecolor=magenta,      
    urlcolor=cyan,
}

\begin{document}

\articletype{Paper} 

\title{State change via one-dimensional scattering in quantum mechanics}

\author{Olivia Pomerenk$^{1,2*}$\orcid{0000-0002-0481-3952} and Charles S. Peskin$^{1}$\orcid{0000-0003-3749-9864}}

\affil{$^1$Courant Institute of Mathematical Sciences, New York University, New York, NY 10012, USA}

\affil{$^2$Present address: School of Engineering, Brown University, Providence, RI 02912, USA}

\affil{$^*$Author to whom any correspondence should be addressed.}

\email{olivia\_pomerenk@brown.edu}


\begin{abstract}
This study aims to address the nature of state change, measurement, and probabilistic outcomes in non-relativistic quantum mechanics.
We consider a pair of particles that interact in a one-dimensional
setting via a delta-function potential. One of the particles is
confined to a one-dimensional box, and the other particle is free.
The free particle is incident from the left with specified energy, and
it may cause changes in state of the confined particle before flying
away to the left or to the right. We present a formulation and computational scheme that avoids the
 use of perturbation theory and determines the probability of any
 such outcome as a function of the initial state of the confined
 particle and the energy of the incident particle. As demonstrated by a direct comparison, this presented method holds multiple advantages over a standard perturbative method. The problem formulation and corresponding computational scheme may have applications in physical settings which admit one-dimensional scattering, e.g., in the study of quantum wires or quantum dots.
\end{abstract}

\section{Introduction}

In ordinary non-relativistic quantum mechanics, which is the setting for the present paper, state change of a system can happen as a result of interaction with another system. The interaction can be modeled by the time-dependent Schr{\"o}dinger equation, which is typically solved by time-dependent perturbation theory \citep{langhoff1972aspects, suzuki1983degenerate, karplus1963variation, primas1963generalized}. The evolution described by the Schr{\"o}dinger equation is continuous and deterministic, but the observed final state of the system is discrete and random \citep{omnes2018interpretation}. This gap between what is computed and what is observed is bridged by the theory of measurement, which does not actually explain the full underlying physics but does give a definite prescription of how the results obtained by solving the Schr{\"o}dinger equation are to be translated into a discrete list of possible outcomes together with their probabilities.

In the present paper we consider a particularly simple example of
state change via scattering, in which the measurement step is
essentially built into the interaction process, and in which we are
able to compute the probabilities of a finite number of different
possible outcomes without resort to any kind of perturbation theory.
This is done by formulating the problem as an infinite system of
integral equations, which are solved numerically with second-order
accuracy. This approach thus differs from that of previous works, e.g., \citep{langhoff1972aspects,soffer1998nonautonomous, choi2020perturbation}, in which time-dependent perturbation theory is used to characterize the effect of an external non-autonomous perturbation on a bound state. This work instead treats the composite system (external system interacting with a bound state) altogether, characterizing the asymptotic states which arise after the interaction has taken place. A major advantage of the method developed in this work is that the presented approach can handle arbitrarily large strengths of interaction. This is a distinct advantage over perturbative methods, which are limited to addressing only sufficiently small interaction strengths. This will be shown explicitly by comparison of our approach against a standard perturbative method.

In one spatial dimension, we consider a two-particle system.  One of
the particles is confined to a box (i.e., an interval) of length $L$,
but the walls of the box are transparent to the other particle, which
moves freely on the whole real line. The two particles interact via a
delta-function potential, which may be attractive or repulsive. This choice of potential is standard in the literature \citep{boos2018quantum,dollard1978time,erman2018scattering,harrison2016quantum} and is frequently chosen for mathematical simplicity.
In the scattering problem considered here, there is an incoming
state, in which the the free particle is incident from the left
with a definite energy, and the confined particle is in a particular stationary state of a particle in a one-dimensional box.
This incoming state is a product state of the wavefunctions
of the two particles, which means that the particles are
independent in the distant past, before any interaction has
occurred.  The outgoing state is \textit{not} a product state,
however.  The interaction has correlated the states of the
two particles, so that observation of the outgoing state of
the free particle determines the corresponding state of the confined
particle (assuming that the incoming states were known).

This setup may be viewed as an idealized analog of a real, quasi-one-dimensional scattering system -- such as quantum wires or quantum dots \citep{vaishnav2006hall, goncharov2014semispectral,jacak2013quantum, kouwenhoven1998quantum,harrison2016quantum} --  which involve particles, such as electrons, interacting with imperfections or internal structures (e.g., alloy scattering \citep{tsetseri2004mobility}, impurity scattering \citep{vaishnav2006hall}, or Fano resonance \citep{kobayashi2003fano}).

Three features in particular may be pointed out as
 idealizations: the one-dimensional setting, the infinite depth of the
 potential well in which one of the particles is confined, and the
 localized nature of the interaction between the two particles.  These
 idealizations are made because our purpose here is to construct the
 simplest case possible in which discrete state change (a ``quantum
 jump'') comes about as a result of an interaction.  For an
 interaction, we need at least two particles, so the configuration
 space needs to be at least two-dimensional.  For discrete state
 change, we need discrete states, so at least one of the particles has
 to be confined.  A finite-depth well could be considered, but in that
 case the configuration space of the system would be the whole
 $x_1,x_2$ plane instead of the strip shown in Fig. \ref{fig:setup}(d), and there
 would be only a finite number of discrete stationary states of the
 partially confined particle.  The discrete states would not then form
 a basis for all possible states of the (imperfectly) confined
 particle, and we would have to consider also the continuum of states
 in which that particle has too much energy to be confined by the
 finite depth well.  These states would be needed mathematically as
 part of a basis for the state space of the system even to describe
 situations in which the system has not enough energy for these
 unbound states to be reached. This is a considerable complication
 that we avoid here considering an infinite-depth wall, as is often
 done as an approximation in applications involving partially (i.e.,
 temporarily) trapped particles \citep{boos2018quantum,dollard1978time,erman2018scattering,harrison2016quantum}.  A more subtle point
 about the choice of system to consider is that one of the two
 particles must be free so that the two particles can be
 independent in the distant past, and yet entangled despite being
 separated in the distant future.  This is an essential feature of
 measurement in quantum mechanics.



\section{Problem statement}

\begin{figure}
  \centerline{\includegraphics[scale=0.8]{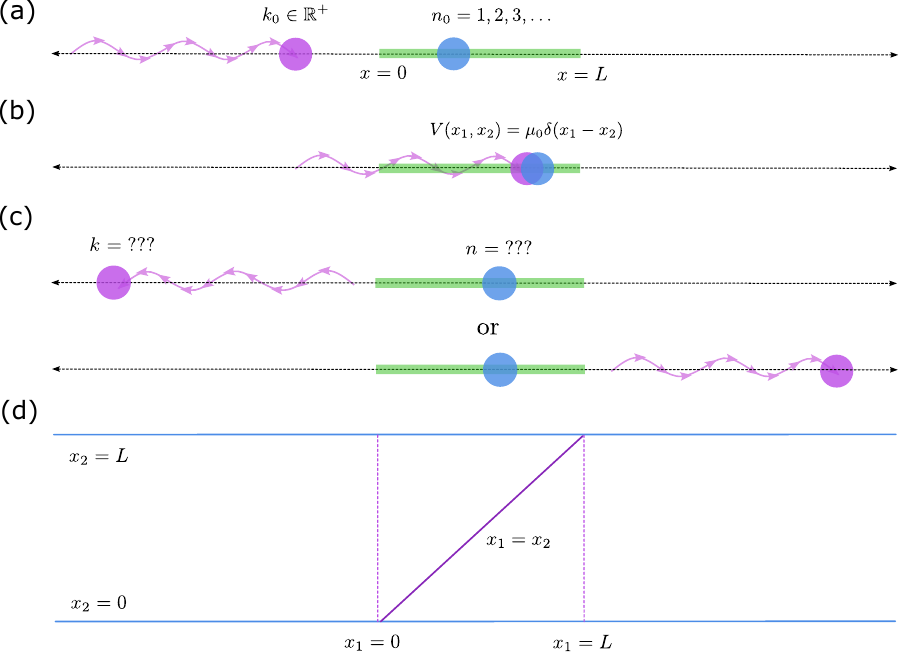}}
  \caption{\label{fig:setup} Schematic of scattering problem considered in the present paper. (a) A free incident particle (pink), associated with wavenumber $k_0$, approaches from the left. A trapped particle (blue) within the interval $[0,L]$ (green) is associated with energy index $n_0$. (b) The particles interact via a delta-function potential which has strength $\mu_0$. (c) Possible outcomes involve the incident particle scattering to the left (top) or the right (bottom) with some resultant wavenumber $k$, and leaving the trapped particle in some energy state $n$. (d) The interaction takes place along the line $x_1=x_2$ in configuration space, where $x_1$ and $x_2$ are respectively the positions of the incident and trapped particles.}
\end{figure}

We consider scattering in one spatial dimension by a local interaction between two particles, with one particle free and the other confined to a one-dimensional (1D) box. The particles respectively have mass $m_1$ and $m_2$. As shown schematically in Fig. \ref{fig:setup}(a), the particle with mass $m_2$ is confined to the interval $(0,L)$, while the particle with mass $m_1$ is free.

The configuration space of the system is 
\begin{equation}
    \mathbb{R}\times (0,L),
\end{equation}
and the time-independent Schr{\"o}dinger equation is
\begin{equation}\label{schrodinger}
    -\frac{\hbar^2}{2m_1}\frac{\partial^2\Psi}{\partial x_1^2} -\frac{\hbar^2}{2m_2}\frac{\partial^2\Psi}{\partial x_2^2} + \mu_0\delta(x_1-x_2)\Psi = E\Psi
\end{equation}
with boundary conditions
\begin{equation}\label{boundary_conditions}
    \Psi(x_1,0) = \Psi(x_1,L) = 0.
\end{equation}
The constant $\mu_0\in\mathbb{R}$ has units of energy $\cdot$ length. Positive $\mu_0>0$ is associated with a repulsive interaction potential, while negative $\mu_0<0$ is associated with an attractive interaction potential. The interaction takes place along the line $x_1=x_2$ in configuration space, as shown in Fig. \ref{fig:setup}(d).


Let
\begin{align}
    \phi_n(x_1) &= \int_0^L\sin\left(\frac{n\pi x_2}{L}\right)\Psi(x_1,x_2)dx_2 \\
    \Psi(x_1,x_2) &= \frac{2}{L}\sum_{n=1}^{\infty}\sin\left(\frac{n\pi x_2}{L}\right)\phi_n(x_1) \label{fourier_decomp}
\end{align}
be a Fourier decomposition of the wavefunction $\Psi$ such that the boundary conditions (\ref{boundary_conditions}) are satisfied by (\ref{fourier_decomp}). To get an equation for $\phi_n(x_1)$, we multiply both sides of (\ref{schrodinger}) by $\sin(n\pi x_2/L)$ and, making use of (\ref{fourier_decomp}), integrate with respect to $x_2$ over $(0,L)$. This yields
\begin{equation}\label{phi_n_equation}
    -\frac{\hbar^2}{2m_1}\frac{d^2\phi_n}{dx^2}(x) + \frac{\hbar^2}{2m_2}\left(\frac{n\pi}{L}\right)^2\phi_n(x) + \frac{2\mu_0}{L}C(x)\sin\left(\frac{n\pi x}{L}\right)\sum_{n'=1}^{\infty}\sin\left(\frac{n'\pi x}{L}\right)\phi_{n'}(x) = E\phi_n(x)
\end{equation}
where we have simplified the notation by writing $x_1=x$, as the integration with respect to $x_2$ has removed all dependencies on $x_2$. Here,
\begin{equation}
    C(x) = \begin{dcases}
        1, \quad x\in(0,L) \\
        0, \quad x\notin (0,L).
    \end{dcases}
\end{equation}
We seek a solution to (\ref{phi_n_equation}) in the form
\begin{equation}\label{phi_n_form}
    \phi_n(x) = \delta_{nn_0}e^{ik_0 x} + \Phi_n(x)
\end{equation}
in which $\Phi_n(x)$ involves only outgoing waves or decays exponentially as $|x|\to \infty$. Here, $n_0$ corresponds to the index of the stationary state of the confined particle in the 1D box, and $k_0$ is the wavenumber of the incident free particle. 

The ansatz (\ref{phi_n_form}) generalizes a standard ansatz made in one-dimensional scattering of a particle by a fixed potential (see for example equation (10.3) of \citep{Tong2017AQM}).  Our problem is more complicated, however, because the scatterer is not a fixed potential but instead is a potential that depends on (and at the same time influences) the position of a confined particle. This particular form (\ref{phi_n_form}) is chosen to represent the composite state of the incoming wave and the confined particle as a product state, as the two sub-systems must be independent. In physical (i.e., non-Fourier) space, this product state takes the form $e^{ik_0 x_1}\sin(n_0\pi x_2/L)$, which has the interpretation that the free particle (with coordinate $x_1$) is in a state of definite momentum and energy with wavenumber $k_0$ and is completely de-localized, while the confined particle (with coordinate $x_2$) is in the stationary state with index $n_0$ of a particle in a 1D box. The Fourier transform with respect to $x_2$ of this product state is $\delta_{nn_0}e^{ik_0 x}$, which is the form given above in (\ref{phi_n_form}). Thus, there is no correlation between the particles in the incoming state, but correlation is generated by the interaction in the outgoing state, as shown schematically in Figs. \ref{fig:setup}(b) and (c).

We require that the term $\delta_{nn_0}e^{ik_0 x}$ be a solution to (\ref{phi_n_equation}) for $x\notin(0,L)$. This implies that the total energy is
\begin{equation}\label{energy}
    E = \frac{\hbar^2}{2m_1}k_0^2 + \frac{\hbar^2}{2m_2}\left(\frac{n_0\pi}{L}\right)^2.
\end{equation}
Then, substituting the ansatz of (\ref{phi_n_form}), multiplying by $2m_1/\hbar^2$, and making use of (\ref{energy}), (\ref{phi_n_equation}) becomes

\begin{equation}\label{big_phi_n_2}
\begin{split}
        -\frac{d^2\Phi_n}{dx^2}(x) &+ \left(\frac{m_1}{m_2}\left(\frac{\pi}{L}\right)^2(n^2-n_0^2)-k_0^2\right)\Phi_n(x)  \\
        &= -\frac{4m_1\mu_0}{\hbar^2 L}C(x)\sin\left(\frac{n\pi x}{L}\right)\sum_{n'=1}^{\infty}\sin\left(\frac{n'\pi x}{L}\right)(\delta_{n'n_0}e^{ik_0 x} + \Phi_{n'}(x)).
\end{split}
\end{equation}
Let 
\begin{equation}\label{an_def}
        c_n = \frac{m_1}{m_2}\left(\frac{\pi}{L}\right)^2(n^2-n_0^2)-k_0^2,\quad\quad n=1,2,\dots 
\end{equation}
Then $c_1<0$ and $c_n>0$ for sufficiently large $n$. Assume that $\nexists n$ such that $c_n=0$. Next, define
\begin{align}
    k_n &= \sqrt{-c_n},\quad\quad c_n < 0 \label{kn_def} \\
    \lambda_n &= \sqrt{c_n},\quad\quad~~ c_n > 0. \label{lambdan_def}
\end{align}
For $n$ such that $c_n<0$, introduce the Green's function
\begin{equation}\label{Gn_k}
    G_n(x) = -\frac{1}{2ik_n}\begin{dcases}
        e^{-ik_n x},\quad &x<0 \\
        e^{ik_n x},\quad &x>0
    \end{dcases}
\end{equation}
and for $n$ such that $c_n>0$, define
\begin{equation}\label{Gn_lambda}
    G_n(x) = \frac{1}{2\lambda_n}\begin{dcases}
        e^{\lambda_n x},\quad &x<0 \\
        e^{-\lambda_n x},\quad &x>0.
    \end{dcases}
\end{equation}
Then, for all $n=1,2,\dots$, $G_n$ satisfies the standard relation \citep{duffy2015green}
\begin{equation}
    -\frac{d^2G_n}{dx^2}(x) + c_nG_n(x) = \delta(x).
\end{equation}
It follows by linearity that (\ref{big_phi_n_2}) may be rewritten as
\begin{equation}\label{pre_22}
    \Phi_n(x)  = -\frac{4m_1\mu_0}{\hbar^2 L}\int_0^LG_n(x-x')\sin\left(\frac{n\pi x'}{L}\right)\sum_{n'=1}^{\infty}\sin\left(\frac{n'\pi x'}{L}\right)(\delta_{n'n_0}e^{ik_0 x'} + \Phi_{n'}(x'))dx'.
\end{equation}
Defining
\begin{equation}\label{Ann}
    A_{nn'}(x) = \frac{4m_1\mu_0}{\hbar^2 L}\sin\left(\frac{n\pi x}{L}\right)\sin\left(\frac{n'\pi x}{L}\right),
\end{equation}
(\ref{pre_22}) becomes
\begin{equation}\label{22}
    \Phi_n(x) + \sum_{n'=1}^\infty \int_0^L G_n(x-x')A_{nn'}(x')\Phi_{n'}(x')dx' = -\int_0^L G_n(x-x')A_{nn_0}(x')e^{ik_0x'}dx'.
\end{equation}
This is a system of Fredholm integral equations of the second kind \citep{tricomi1985integral}. These equations may be solved in a traditional manner using perturbation theory in the weak-interaction limit (Appendix A), or we may proceed numerically to solve these equations without any restriction on the strength of the interaction.

\section{Evaluation of the outcome probabilities}

Assuming that we can solve (\ref{22}), we discuss in this section how the solution can be used to evaluate the probability of each possible outcome of the interaction between the free and the bound particle. This requires the notion of probability flux, which is derived by considering the time-dependent Schr{\"o}dinger equation,
\begin{equation}\label{time_dependent_SE}
    i\hbar\frac{\partial\psi}{\partial t} = -\frac{\hbar^2}{2m_1}\frac{\partial^2\psi}{\partial x_1^2} -\frac{\hbar^2}{2m_2}\frac{\partial^2\psi}{\partial x_2^2} + \mu_0\delta(x_1-x_2)\psi.
\end{equation}
Multiplying both sides of (\ref{time_dependent_SE}) by the complex conjugate of $\psi$, taking the conjugate of both sides of the resulting equation, and subtracting this conjugated equation from the non-conjugated equation yields
\begin{equation}
    \frac{\partial}{\partial t}|\psi|^2 + \frac{\partial}{\partial x_1}\frac{\hbar}{2im_1}\left(\overline{\psi}\frac{\partial \psi}{\partial x_1} - \psi\frac{\partial \overline{\psi}}{\partial x_1}\right) + \frac{\partial}{\partial x_2}\frac{\hbar}{2im_2}\left(\overline{\psi}\frac{\partial \psi}{\partial x_2} - \psi\frac{\partial \overline{\psi}}{\partial x_2}\right) = 0,
\end{equation}
where $\overline{\psi}$ denotes the complex conjugate of $\psi$. This has the form of a continuity equation, i.e.,
\begin{equation}
    \frac{\partial\rho}{\partial t} + \frac{\partial f_1}{\partial x_1} + \frac{\partial f_2}{\partial x_2} = 0
\end{equation}
where
\begin{align}
    \rho &= |\psi|^2 \label{23} \\
    f_1 &= \frac{\hbar}{2im_1}\left(\overline{\psi}\frac{\partial \psi}{\partial x_1} - \psi\frac{\partial \overline{\psi}}{\partial x_1}\right) \label{f1_psi} \\
    f_2 &= \frac{\hbar}{2im_2}\left(\overline{\psi}\frac{\partial \psi}{\partial x_2} - \psi\frac{\partial \overline{\psi}}{\partial x_2}\right) \label{25}.
\end{align}
Thus, $\rho$ is a probability density, and $f_1$ and $f_2$ are fluxes associated respectively with particles 1 and 2. The transition from the time-dependent to the time-independent
 Schrodinger equation is made by assuming that $\psi(x_1,x_2,t) = \Psi(x_1,x_2)\exp(-i(E/\hbar)t)$, where $E$ is the energy and $\hbar$ is
 Planck's constant. Note that $\Psi(x_1,x_2)$ is not necessarily real. Indeed, if it is real, then there is no probability flux. The factor
 $\exp(-i(E/\hbar)t)$ cancels out on the right-hand sides of (\ref{23}-\ref{25}) because in every term it is multiplied by its conjugate.  Thus
 equations (\ref{23}-\ref{25}) hold also with $\psi$ replaced by $\Psi$:
 \begin{align}
    \rho &= |\Psi|^2  \\
    f_1 &= \frac{\hbar}{2im_1}\left(\overline{\Psi}\frac{\partial \Psi}{\partial x_1} - \Psi\frac{\partial \overline{\Psi}}{\partial x_1}\right) \label{f1} \\
    f_2 &= \frac{\hbar}{2im_2}\left(\overline{\Psi}\frac{\partial \Psi}{\partial x_2} - \Psi\frac{\partial \overline{\Psi}}{\partial x_2}\right).
\end{align}

To obtain probabilities for all possible outcomes associated with particle 1, define the integral of the flux for particle 1 over all possible positions of particle 2,
\begin{equation}
    F(x_1) = \int_0^Lf_1(x_1,x_2)dx_2.
\end{equation}
To evaluate $F(x_1)$, we make use of (\ref{fourier_decomp}) and (\ref{f1}) to obtain
\begin{equation}\label{40}
    \overline{\Psi}\frac{\partial \Psi}{\partial x_1} = \left(\frac{2}{L}\right)^2\sum_{n,n'=1}\sin\left(\frac{n\pi x_2}{L}\right)\sin\left(\frac{n'\pi x_2}{L}\right)\overline{\phi_{n'}}(x_1)\frac{\partial \phi_n}{\partial x_1}(x_1).
\end{equation}
Integrating both sides of (\ref{40}) over $(0,L)$ with respect to $x_2$ removes the dependence on $x_2$, and so we again drop the subscript 1 on $x$ for ease of notation. We obtain
\begin{equation}\label{41}
    F(x) = \frac{\hbar}{2im_1}\frac{2}{L}\sum_{n=1}^{\infty}\left(\overline{\phi_n}(x)\frac{d\phi_n}{dx}(x) - \phi_n(x)\frac{d\overline{\phi_n}}{dx}(x)\right).
\end{equation}
Note the absence of interference terms between different $n$.

For $x>L$, we write
\begin{equation}\label{42}
    \phi_n(x) = \begin{dcases}
        \phi_n(L)e^{ik_n(x-L)}\quad~ c_n < 0 \\
        \phi_n(L)e^{-\lambda_n(x-L)}\quad c_n > 0.
    \end{dcases}
\end{equation}
Then, for $n$ such that $c_n>0$,
\begin{equation}
    \overline{\phi_n}(x)\frac{d\phi_n}{dx}(x) - \phi_n(x)\frac{d\overline{\phi_n}}{dx}(x) = 0.
\end{equation}
Substitution of (\ref{42}) into (\ref{41}) therefore implies that, for $x>L$,
\begin{equation}\label{Fxright}
    F(x) = \frac{2}{L}\sum_{n:c_n<0}\frac{\hbar k_n}{m_1}\rvert\phi_n(L)|^2 = \frac{2}{L}\sum_{n:c_n<0}\frac{\hbar k_n}{m_1}\big\rvert\Phi_n(L)+\delta_{nn_0}e^{ik_0L}\big|^2.
\end{equation}

For $x<0$ and $n\neq n_0$, a similar argument holds with $k_n\to -k_n$, $\lambda_n\to-\lambda_n$, and $\phi_n(L)\to\phi_n(0)$. For the particular case of $n=n_0$ and $x<0$, we have
\begin{equation}
    \phi_{n_0}(x) = e^{ik_0x} + \Phi_{n_0}(0)e^{-ik_0x}
\end{equation}
and so
\begin{equation}\label{Fxleft}
    F(x) = \frac{2}{L}\frac{\hbar k_0}{m_1}-\frac{2}{L}\sum_{n:c_n<0}\frac{\hbar k_n}{m_1}\big\rvert\Phi_n(0)\big\rvert^2.
\end{equation}
Note that $F(x)$ is independent of $x$, which implies conservation of probability.

We see that $2\hbar k_0/Lm_1$ is the incoming flux of probability, so we divide $F(x)$ by this term and make the definitions
\begin{align}
    p_n^+ &= \frac{k_n}{k_0}\big\rvert \Phi_n(L)+\delta_{nn_0}e^{ik_0L}\big\rvert^2 ,\label{pnplus} \\
    p_n^- &= \frac{k_n}{k_0}\big\rvert \Phi_n(0)\big\rvert^2 , \label{pnminus}
\end{align}
for $n$ such that $c_n<0$. Here, $p_n^+$ is the probability that particle 1 exits to the right with wavenumber $k_n$, leaving particle 2 in the $n^{\text{th}}$ stationary state of a particle in a box, and $p_n^-$ is the probability that particle 1 exits to the left, leaving particle 2 in the $n^{\text{th}}$ stationary state. The total probability that particle 2 is in stationary state $n$ as a consequence of the interaction is thus
\begin{equation}
    p_n = p_n^+ + p_n^-.
\end{equation}

\subsection{Restriction to finitely many outcomes}
There are a finite number of possible outcomes of (\ref{pnplus}) and (\ref{pnminus}) because of the restriction that $c_n<0$. Each of the different outcomes conserves energy. The energy of the incoming state, by (\ref{energy}), is
\begin{equation}
    E_0 = \frac{\hbar^2}{2m_1}k_0^2 + \frac{\hbar^2}{2m_2}\left(\frac{n_0\pi}{L}\right)^2.
\end{equation}
The energy for either of the states $n$ (i.e., with particle 1 exiting to the right or left while leaving particle 2 in the $n^{\text{th}}$ stationary state) is
\begin{equation}\label{energy_outcome}
    E_n = \frac{\hbar^2}{2m_1}k_n^2 + \frac{\hbar^2}{2m_2}\left(\frac{n\pi}{L}\right)^2.
\end{equation}
Substituting (\ref{an_def}) and (\ref{kn_def}) into the above yields $E_n=E_0$.

Further, it must be the case that
\begin{equation}\label{sum_to_1}
    \sum_{n:c_n<0}(p_n^+ + p_n^-) = \sum_{n:c_n<0}p_n= 1.
\end{equation}
That is, only those outcomes $n$ such that $c_n<0$ are possible, and indeed they represent exactly all of the outcomes with nonzero probability. This is an immediate consequence of the fact that $F(x)$ is independent of $x$: equating the right hands sides of (\ref{Fxright}) and (\ref{Fxleft}) gives an equation which may be manipulated to produce (\ref{sum_to_1}).

In the next section, we present a method with which to solve (\ref{22}) numerically for $\Phi_n(x)$ on the interval $[0,L]$, and to thus compute the transmission and reflection probabilities given by (\ref{pnplus}) and (\ref{pnminus}).

\section{Numerical method}
In this section we describe a numerical method for the solution of (\ref{22}) with arbitrary interaction strength $\mu_0$.  For comparison, in Appendix A, we derive an approximate solution known as the Born approximation, which is valid in the regime of weak interaction, i.e., when $\mu_0$ is small in an appropriate dimensionless sense. We solve (\ref{22}) for $\Phi_n(x)$ with $n=1,...,n_{\text{T}}$ for finite truncation order $n_{\text{T}}$ on the interval $x\in[0,L]$. First, denote the right hand side of (\ref{22}) by
\begin{equation}
    f_n(x) = -\int_0^L G_n(x-x')A_{nn_0}(x')e^{ik_0x'}dx',
\end{equation}
which may be computed explicitly for $x\in[0,L]$. Equation (\ref{22}) may be written as a system of $n_{\text{T}}$ equations:
\begin{equation}
\begin{split}
        &\Phi_1(x) + \sum_{n'=1}^{\infty}\int G_1(x-x')A_{1n'}(x')\Phi_{n'}(x')dx' = f_1(x) \\
        &\Phi_2(x) + \sum_{n'=1}^{\infty}\int G_2(x-x')A_{2n'}(x')\Phi_{n'}(x')dx' = f_2(x)\\
        &\quad\quad\vdots \\
        &\Phi_{n_{\text{T}}}(x) + \sum_{n'=1}^{\infty}\int G_{n_{\text{T}}}(x-x')A_{n_{\text{T}}n'}(x')\Phi_{n'}(x')dx' = f_{n_{\text{T}}}(x).
\end{split}
\end{equation}
In order to make this a system of $n_{\text{T}}$ equations for $n_{\text{T}}$ unknowns, and thus uniquely determine $\Phi_n(x)$, we must truncate the infinite series within each equation at $n'=n_{\text{T}}$:
\begin{equation}\label{continuous_system}
\begin{split}
        &\Phi_1(x) + \sum_{n'=1}^{n_{\text{T}}}\int G_1(x-x')A_{1n'}(x')\Phi_{n'}(x')dx' = f_1(x) \\
        &\Phi_2(x) + \sum_{n'=1}^{n_{\text{T}}}\int G_2(x-x')A_{2n'}(x')\Phi_{n'}(x')dx' = f_2(x)\\
        &\quad\quad\vdots \\
        &\Phi_{n_{\text{T}}}(x) + \sum_{n'=1}^{T}\int G_{n_{\text{T}}}(x-x')A_{n_{\text{T}}n'}(x')\Phi_{n'}(x')dx' = f_{n_{\text{T}}}(x).
\end{split}
\end{equation}
So long as $n_{\text{T}}$ is sufficiently large, this truncation should be appropriate for $x\in[0,1]$. This is because, by (\ref{Gn_k}) and (\ref{Gn_lambda}), the quantity $G_n(x)$ decays exponentially as $n\to\infty$ for any $x\in\mathbb{R}$. Thus, $|\Phi_n(x)|\to 0$ for all $x$.

Now we discretize equations (\ref{continuous_system}) using $N$ equally spaced quadrature nodes $\lbrace x_1, ..., x_N\rbrace$ on $[0,L]$ such that $x_1=0$ and $x_N=L$. Introduce the notation for the kernel, which may be computed explicitly for any $(n_1,n_2,x_i,x_j)$:
\begin{equation}
    K_{n_1n_2}^{ij} = G_{n_1}(x_i-x_j)A_{n_1n_2}(x_j).
\end{equation}
Then, a discretization of (\ref{continuous_system}) is
\begin{equation}\label{discretized_system}
    \begin{split}
        \sum_j \left[(\delta_{ij} + K_{11}^{ij}w_j)\Phi_1(x_j) + K_{12}^{ij}\Phi_2(x_j)w_j + \dots + K_{1n_{\text{T}}}^{ij}\Phi_{n_{\text{T}}}(x_j)w_j\right] &= f_1(x_i) \\
        \sum_j \left[K_{21}^{ij}\Phi_1(x_j)w_j + (\delta_{ij}+K_{22}^{ij}w_j)\Phi_2(x_j) + \dots + K_{2n_{\text{T}}}^{ij}\Phi_{n_{\text{T}}}(x_j)w_j\right] &= f_2(x_i) \\
        &\vdots \\
        \sum_j\left[ K_{n_{\text{T}}1}^{ij}\Phi_1(x_j)w_j +K_{n_{\text{T}}2}^{ij}\Phi_2(x_j)w_j + \dots + (\delta_{ij} + K_{n_{\text{T}}n_{\text{T}}}^{ij}w_j)\Phi_{n_{\text{T}}}(x_j)\right] &= f_{n_{\text{T}}}(x_i)
    \end{split}
\end{equation}
where the $w_j$ are precomputed quadrature weights for the nodes $x_j$. Here, we use trapezoidal quadrature. Each line of the above discretization corresponds to $N$ equations, one for each of the $N$ quadrature nodes. As there are $n_{\text{T}}$ lines of (\ref{discretized_system}), we have a system of $Nn_{\text{T}}$ equations for $Nn_{\text{T}}$ unknowns.

To solve this numerically, we construct a square block matrix $B$ which has size $Nn_{\text{T}}\times Nn_{\text{T}}$. There are $n_{\text{T}}^2$ blocks, each of which contains an $N\times N$ matrix. At the $(n_1,n_2)$ block, the $(i,j)$ entry is
\begin{equation}
    B^{ij} = \delta_{ij}\delta_{n_1n_2} + K_{n_1n_2}^{ij}w_j.
\end{equation}
Note that $n_1,n_2=1,...,n_{\text{T}}$ and $i,j=1,...,N$. The system may then be expressed in the block matrix form
\begin{equation}\label{matrix_system}
\bm{B}\Phi = 
    \left[I + \begin{pmatrix}
        \mathbf{K}_{11} & \dots & \mathbf{K}_{1n_{\text{T}}} \\
        \vdots & \ddots & \vdots \\
        \mathbf{K}_{n_{\text{T}}1} & \dots & \mathbf{K}_{n_{\text{T}}n_{\text{T}}}
    \end{pmatrix} \begin{pmatrix}
        \mathbf{w} & \dots & 0 \\
        \vdots & \ddots & \vdots \\
        0 & \dots & \mathbf{w}
    \end{pmatrix}\right]\Phi = f
\end{equation}
where
\begin{equation}
    \mathbf{w} = \begin{pmatrix}
        w_1 & \dots & 0\\
        \vdots & \ddots & \vdots \\
        0 & \dots & w_N
    \end{pmatrix}
\end{equation}
and
\begin{equation}
    \mathbf{K}_{n_1n_2} = \begin{pmatrix}
        K_{n_1n_2}^{11} & \dots & K_{n_1n_2}^{1N}\\
        \vdots & \ddots & \vdots \\
        K_{n_1n_2}^{N1} & \dots & K_{n_1n_2}^{NN}
    \end{pmatrix}.
\end{equation}
The solution $\Phi_n(x_i)$ is computed by numerically solving (\ref{matrix_system}), e.g., by using the backslash operator in MATLAB \citep{kattan2011solving}. With this solution in hand, we compute state change probabilities of the confined particle via (\ref{pnplus}) and (\ref{pnminus}).

\section{Calculation of state change probabilities}

Here we present the structure of state change probabilities in a variety of cases. Throughout the section, we take the special case $m_1=m_2:=m$, i.e., both particles have equal mass. Although the formulation and corresponding numerical method can handle arbitrary $m_1$ and $m_2$, we take the special case of equal masses here to give illustrative results. In our code which implements the numerical method, $m_1$ and $m_2$ are arbitrary parameters. This code is publicly available in a Github repository (see Data Availability statement), so the interested reader may explore the consequences of unequal particle masses.


\subsection{Dimensionless quantities}
Here we introduce a set of dimensionless quantities with which to characterize this problem. These will guide the numerical experiments that follow. The characteristic unit of mass is $m$, the unit of length is $L$, and the unit of time is $T=mL^2/\hbar$, so that $\hbar=1$ in these units. Then, a unit of energy is $E_U = \hbar/T=\hbar^2/(mL^2)$.

The dimensionless energy of the incident particle is then
\begin{equation}
    \tilde{E}=\frac{E}{E_U}=\frac{mL^2E}{\hbar^2} = L^2k_0^2 
\end{equation}
 and the dimensionless interaction strength is
\begin{equation}
    g = \frac{mL\mu_0}{\hbar^2}.
\end{equation}
The third dimensionless parameter is $n_0$, the stationary state of the confined particle in the incoming state.

We may then define the dimensionless quantity
\begin{equation}\label{epsilon}
    \epsilon = \frac{g}{\sqrt{\tilde{E}}} = \frac{g}{Lk_0}=\frac{m\mu_0}{k_0\hbar^2},
\end{equation}
which encodes the strength of the interaction relative to the dimensionless wavenumber of the incident particle. This quantity will be useful to characterize the numerical experiments that follow.

We will explore the effects of varying the dimensionless interaction strength $g$, the dimensionless wavenumber $Lk_0$, and the initial stationary state $n_0$ on the probabilities of change of state of the confined particle. Unless stated otherwise, we use $n_{\text{T}}=50$ truncation terms and $N=2n_{\text{T}}$ grid points in a given numerical discretization.

In the tableaux that follow, we plot all possible outcome probabilities $p_n$ (total probability that the interaction leaves particle 2 in eigenstate $n$), $p_n^-$ (probability that the interaction leaves particle 2 in eigenstate $n$ and particle 1 exits to the left), and $p_n^+$ (probability that the interaction leaves particle 2 in eigenstate $n$ and particle 1 exits to the right) against incident dimensionless wavenumbers $Lk_0\in(0,30)$ of particle 1. We use 500 evenly spaced values of $Lk_0$ within this interval, with each value of $Lk_0$ corresponding to a distinct numerical solve of (\ref{matrix_system}). Each figure comprises six panels, all of which involve the confined particle lying in a particular energy state $n_0$ prior to the interaction. Within each figure, panels (a-c) are associated with $\epsilon>0$, i.e., a repulsive potential, while panels (d-f) are associated with $\epsilon<0$, i.e., the equal and opposite attractive potential. In each panel, all possible outcomes are shown except for those associated with $n=n_0$, i.e., we only display outcomes which involve a change in the confined particle's energy state. As discussed earlier, there are only finitely many outcomes available for a given value of $Lk_0$.

\subsection{Probability structure in the case $n_0=1$}

\begin{figure}
  \centerline{\includegraphics[width=\textwidth]{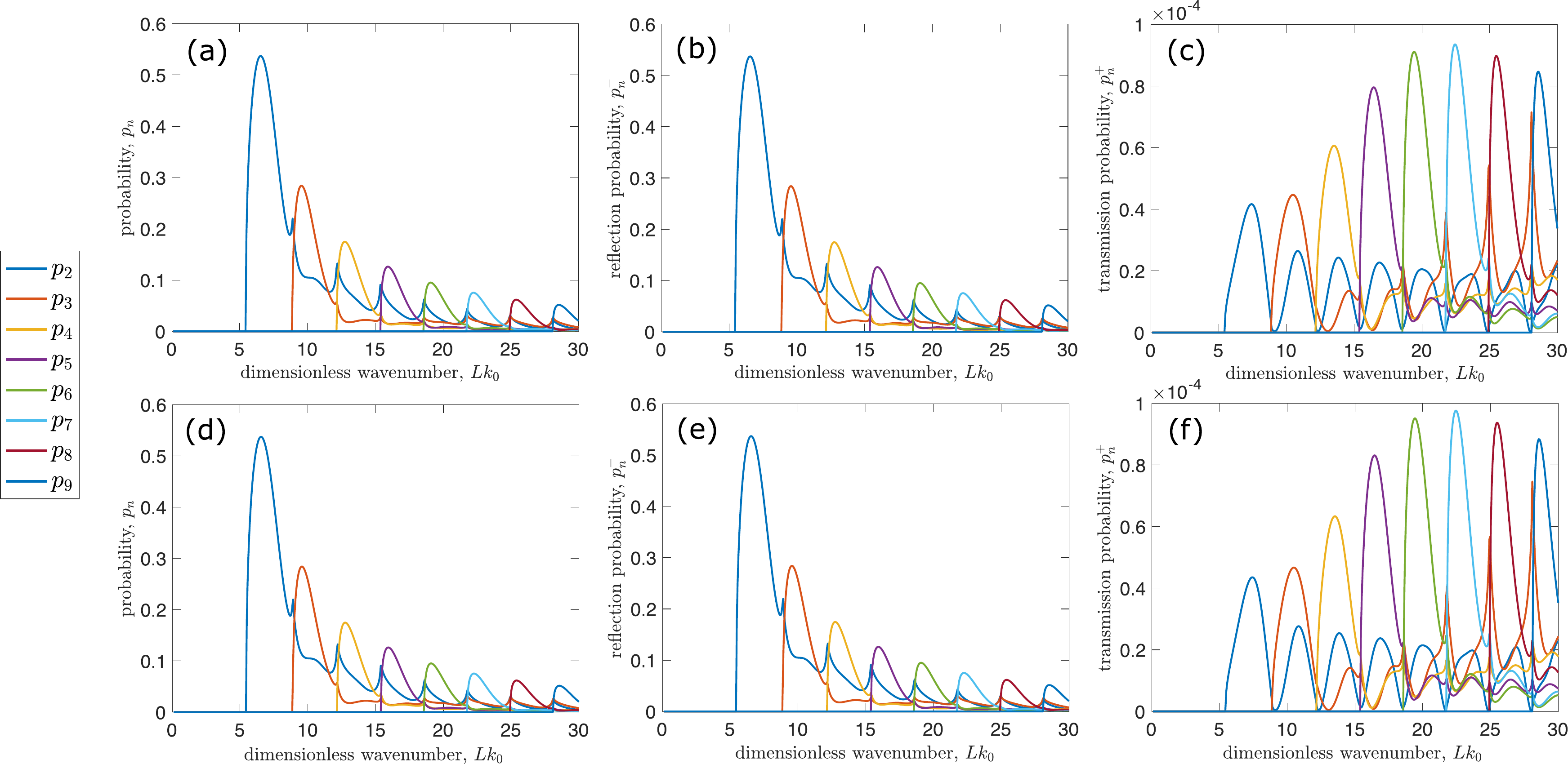}}
  \caption{\label{fig:high_strength_n1} Probability structure with $n_0=1$ and a high interaction strength. The parameter $\epsilon$ ranges from $\mathcal{O}(10^3)-\mathcal{O}(10^5)$. (a-c) are associated with repulsive interaction, and (d-f) are associated with attractive interaction. (a,d) represent total probabilities indexed by the outcome $n$; (b,e) represent reflection probabilities indexed by $n$; and (c,f) represent transmission probabilities indexed by $n$.}
\end{figure}

\begin{figure}
  \centerline{\includegraphics[width=\textwidth]{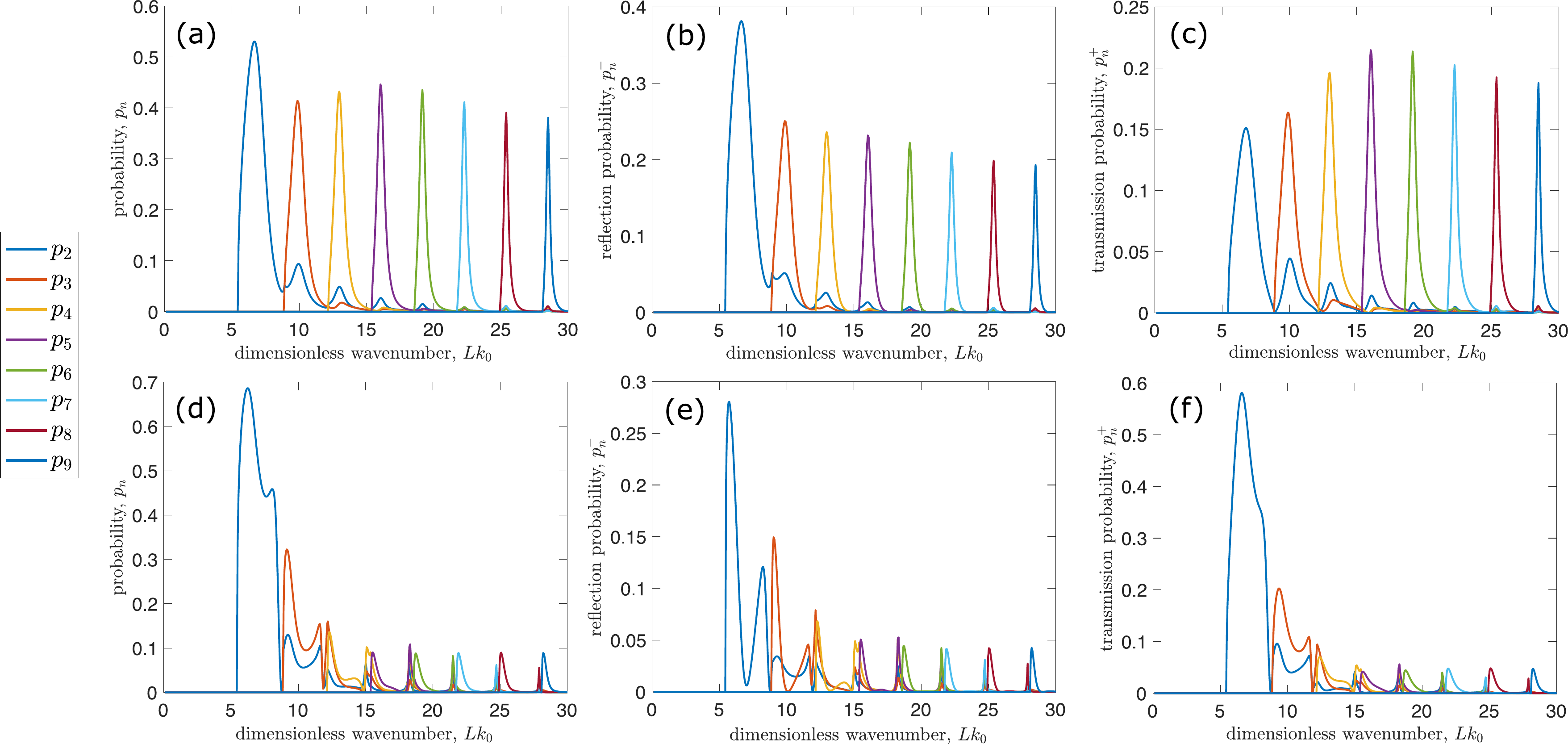}}
  \caption{\label{fig:mid_strength_n1} Probability structure with $n_0=1$ and a moderate interaction strength. The parameter $\epsilon$ ranges from $\mathcal{O}(0.1)-\mathcal{O}(10)$. (a-c) are associated with repulsive interaction, and (d-f) are associated with attractive interaction. (a,d) represent total probabilities indexed by the outcome $n$; (b,e) represent reflection probabilities indexed by $n$; and (c,f) represent transmission probabilities indexed by $n$.}
\end{figure}

\begin{figure}
  \centerline{\includegraphics[width=\textwidth]{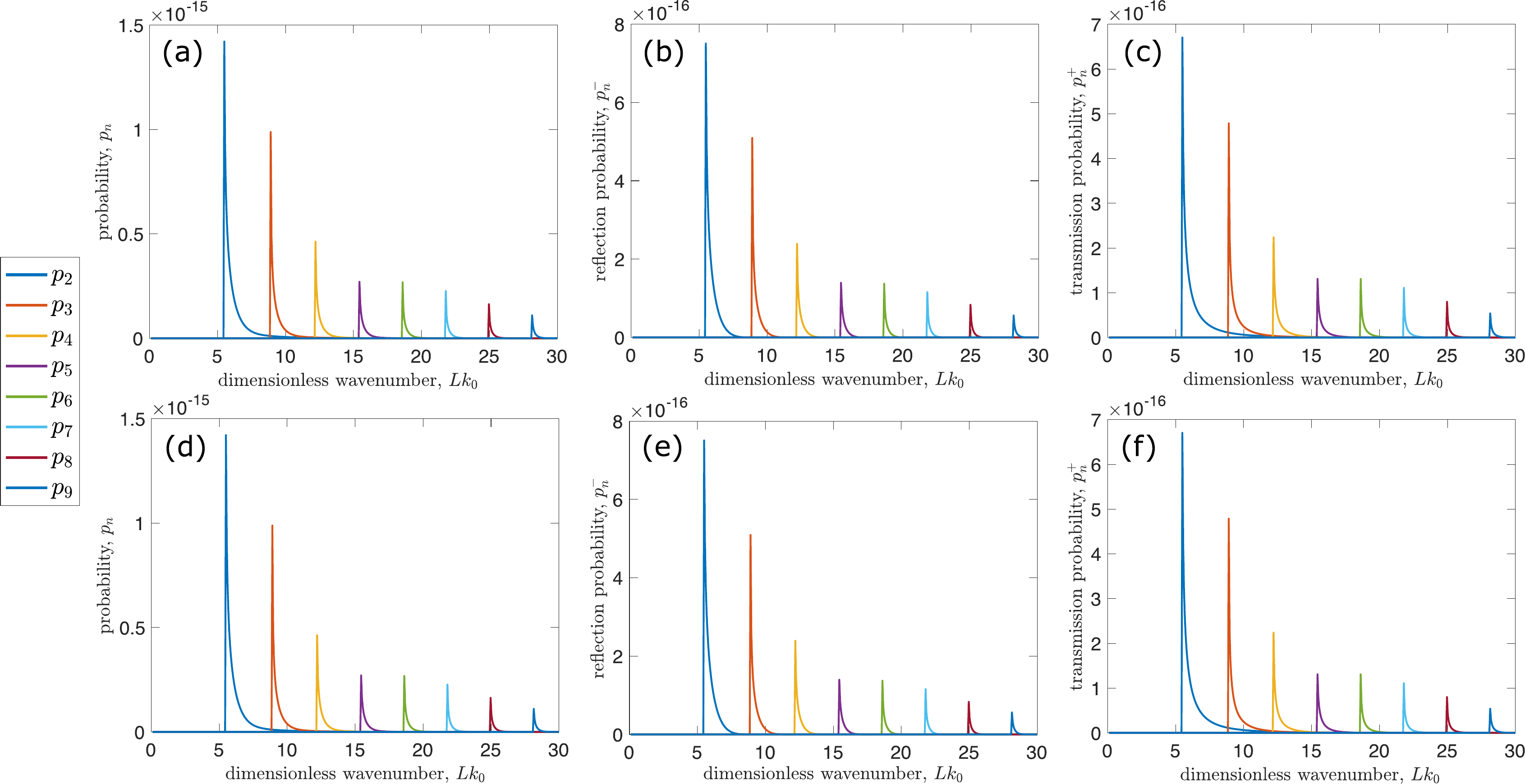}}
  \caption{\label{fig:low_strength_n1} Probability structure with $n_0=1$ and a weak relative interaction strength. The parameter $\epsilon$ ranges from $\mathcal{O}(10^{-9})-\mathcal{O}(10^{-7})$. (a-c) are associated with repulsive interaction, and (d-f) are associated with attractive interaction. (a,d) represent total probabilities indexed by the outcome $n$; (b,e) represent reflection probabilities indexed by $n$; and (c,f) represent transmission probabilities indexed by $n$.}
\end{figure}

We assign the confined particle to be in the ground state prior to the interaction -- that is, we fix $n_0=1$. We show the structure of available outcomes for high (Fig. \ref{fig:high_strength_n1}), moderate (Fig. \ref{fig:mid_strength_n1}), and low (Fig. \ref{fig:low_strength_n1}) interaction strengths.

Fig. \ref{fig:high_strength_n1} displays results for a high interaction strength with $g\sim \mathcal{O}(10^4)$. Correspondingly, the parameter $\epsilon$, which encodes the relative interaction strength, is quite high, and ranges from $\epsilon\sim \mathcal{O}(10^{3})$ to $\epsilon\sim\mathcal{O}(10^5)$. 

We report several comments and observations regarding Fig. \ref{fig:high_strength_n1}. First, further increasing the strength of the interaction (by increasing $g$) even by several orders of magnitude does not appreciably affect the results of Fig. \ref{fig:high_strength_n1}: these represent the limiting high-strength interaction case. Also, perhaps unsurprisingly due to the high interaction strength, the overall probability of reflection (panels b,e) is several orders of magnitude larger than that of transmission (panels c,f). Next, we note that the number of available outcomes, indexed by $n$, increases for higher $Lk_0$. These outcomes grow rapidly yet continuously. Finally, and less intuitively, the structure of outcome probabilities is identical in the repulsive case (panels a-c) as in the attractive case (panels d-f).

Fig. \ref{fig:mid_strength_n1} displays results in the same format as Fig. \ref{fig:high_strength_n1}, with all parameters identical except a decreased $g\sim\mathcal{O}(1)$, which corresponds to a moderate interaction strength. This is associated with $\epsilon\sim\mathcal{O}(0.1)$ to $\epsilon\sim\mathcal{O}(10)$.

Again, we report some observations. In stark contrast to Fig. \ref{fig:high_strength_n1}, Fig. \ref{fig:mid_strength_n1} displays very different outcomes for a repulsive (panels a-c) versus an attractive (panels d-f) interaction potential. In particular, the attractive potential gives rise to a highly disordered array of outcomes, especially for $Lk_0\in(7,20)$. It may be that such choppy behavior is associated with quasi-bound states, in which the incident particle is likely to be trapped but still ultimately exits the box with probability 1. Also, the probability of reflection exceeds that of transmission in the repulsive case, but the opposite is true in the attractive case. Finally, compared to the peaks of Fig. \ref{fig:high_strength_n1}, the peaks of Fig. \ref{fig:mid_strength_n1} are markedly sharper and narrower (although still continuous).

Fig. \ref{fig:low_strength_n1} displays results in the same format as Figs. \ref{fig:high_strength_n1} and \ref{fig:mid_strength_n1}, with all parameters identical except a further decreased $g\sim\mathcal{O}(10^{-8})$, i.e., a low interaction strength. This is associated with $\epsilon\sim\mathcal{O}(10^{-9})$ to $\epsilon\sim\mathcal{O}(10^{-7})$. Just as in the high interaction strength case, we report that the very low interaction results are essentially unchanged (up to the scaling factor $\epsilon^2$) as the interaction strength is decreased further. Also similarly to the high interaction strength case (Fig. \ref{fig:high_strength_n1}), the attractive and repulsive results are identical; these only differ in the moderate interaction strength case (Fig. \ref{fig:mid_strength_n1}). Finally, we note that the peaks of probability display a highly regular pattern, with a peak for a given outcome $n$ arising sharply and then decaying rapidly in a narrow range of $Lk_0$.

\subsection{Probability structure in the case $n_0=5$}

\begin{figure}
  \centerline{\includegraphics[width=\textwidth]{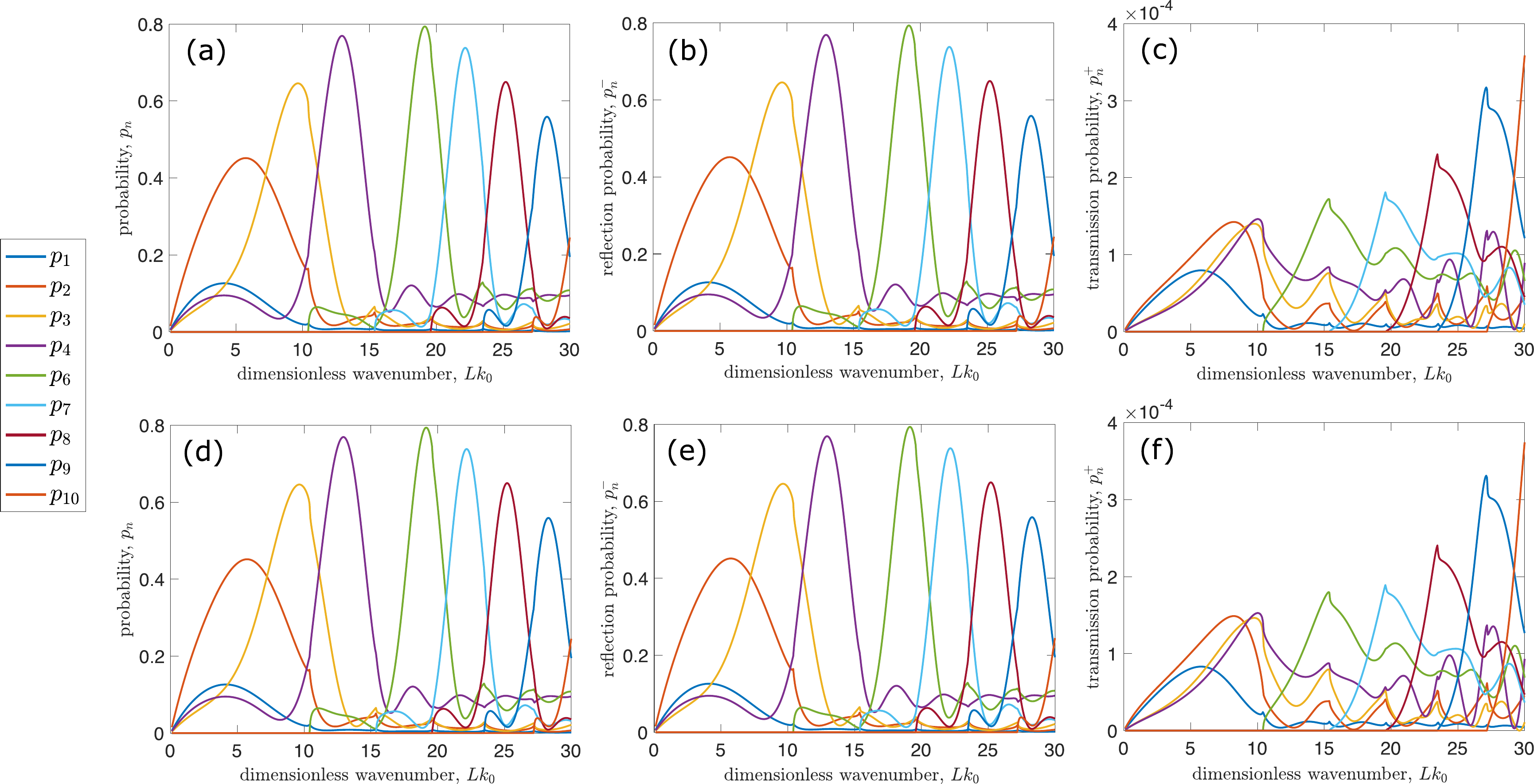}}
  \caption{\label{fig:high_strength_n5} Probability structure with $n_0=5$ and a high interaction strength. The parameter $\epsilon$ ranges from $\mathcal{O}(10^3)-\mathcal{O}(10^5)$. (a-c) are associated with repulsive interaction, and (d-f) are associated with attractive interaction. (a,d) represent total probabilities indexed by the outcome $n$; (b,e) represent reflection probabilities indexed by $n$; and (c,f) represent transmission probabilities indexed by $n$.}
\end{figure}

\begin{figure}
  \centerline{\includegraphics[width=\textwidth]{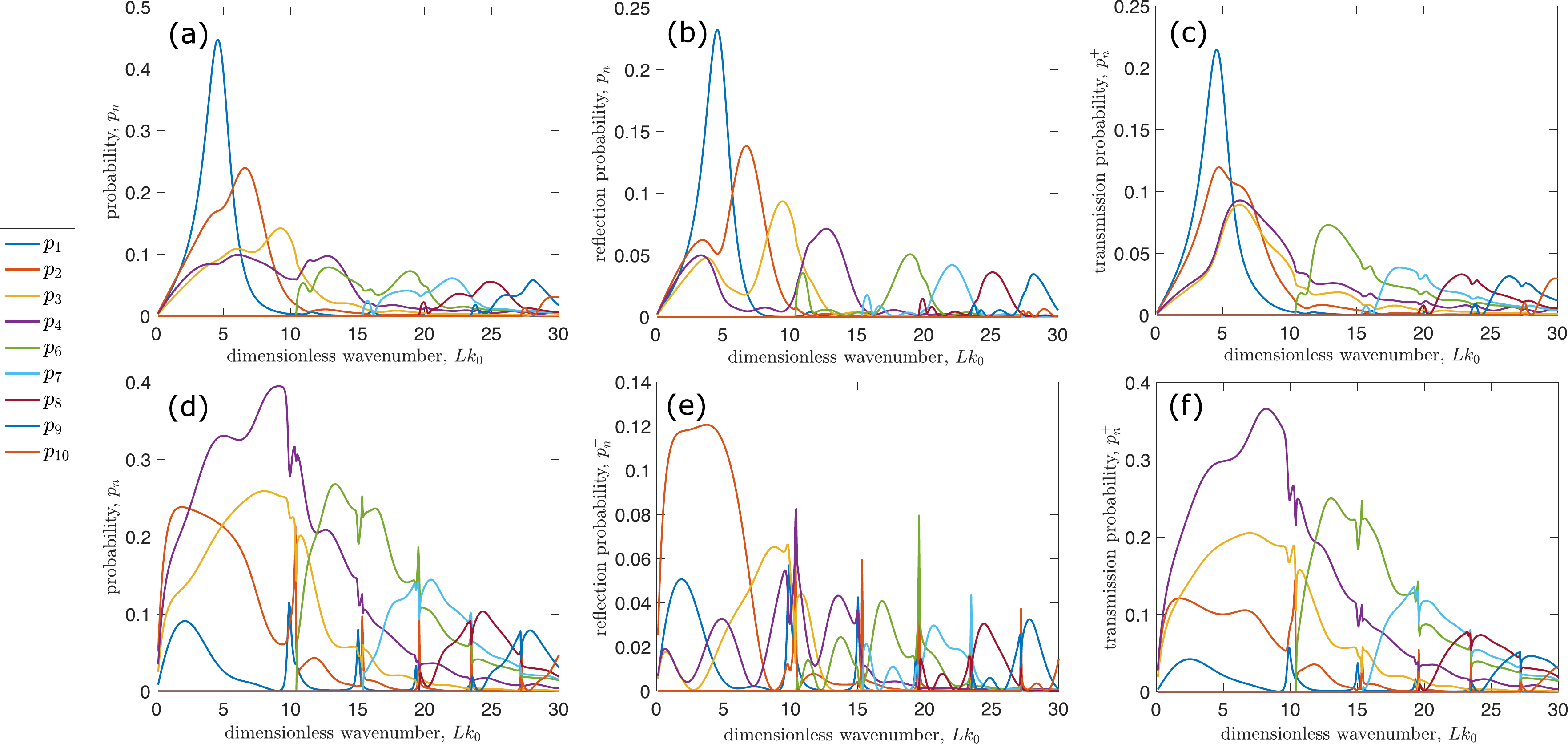}}
  \caption{\label{fig:mod_strength_n5} Probability structure with $n_0=5$ and a moderate interaction strength. The parameter $\epsilon$ ranges from $\mathcal{O}(0.1)-\mathcal{O}(10)$. (a-c) are associated with repulsive interaction, and (d-f) are associated with attractive interaction. (a,d) represent total probabilities indexed by the outcome $n$; (b,e) represent reflection probabilities indexed by $n$; and (c,f) represent transmission probabilities indexed by $n$.}
\end{figure}

\begin{figure}
  \centerline{\includegraphics[width=\textwidth]{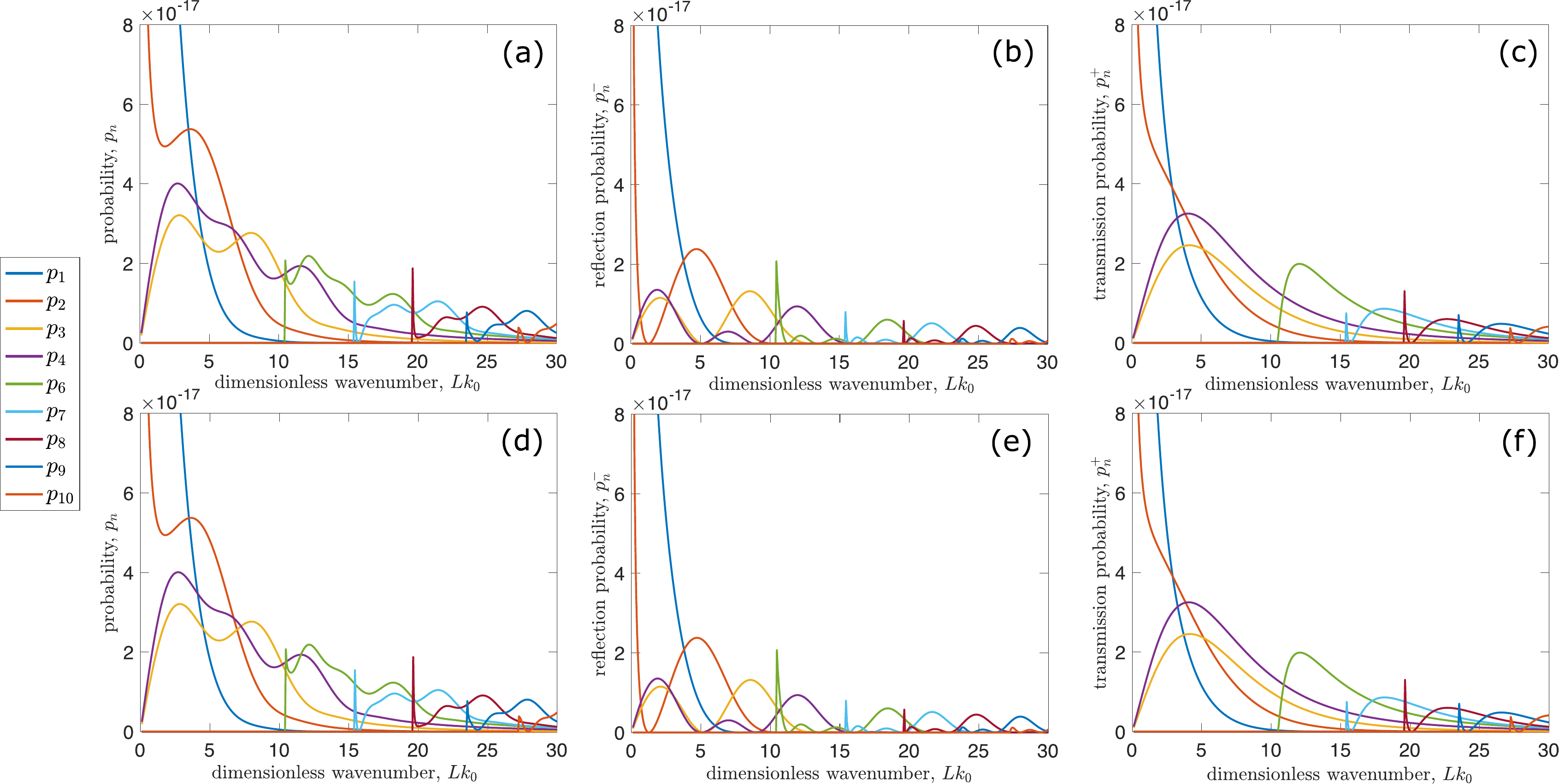}}
  \caption{\label{fig:low_strength_n5} Probability structure with $n_0=5$ and a low interaction strength. The parameter $\epsilon$ ranges from $\mathcal{O}(0.1)-\mathcal{O}(10)$. (a-c) are associated with repulsive interaction, and (d-f) are associated with attractive interaction. (a,d) represent total probabilities indexed by the outcome $n$; (b,e) represent reflection probabilities indexed by $n$; and (c,f) represent transmission probabilities indexed by $n$.}
\end{figure}

We now assign the confined particle to be in a higher-energy stationary state prior to the interaction -- that is, we fix $n_0=5$. We repeat the numerical investigations of the previous section and show results for high (Fig. \ref{fig:high_strength_n5}), moderate (Fig. \ref{fig:mod_strength_n5}), and low (Fig. \ref{fig:low_strength_n5}) interaction potentials.

The results for $n_0=5$ largely mirror those of $n_0=1$. That is, for very high interaction strengths, i.e., large $\epsilon$ (Fig. \ref{fig:high_strength_n5}), as well as for very weak interaction strengths, i.e., small $\epsilon$ (Fig. \ref{fig:low_strength_n5}), the probability structure is identical in the repulsive and attractive cases. Only the moderate-$\epsilon$ case (Fig. \ref{fig:mod_strength_n5}) exhibits different outcomes for attractive and repulsive interactions. Another similarity between the cases $n_0=1$ and $n_0=5$ is that very strong potentials render the probability of transmission to be several orders of magnitude lower than that of reflection -- no matter the state of the confined particle, the incident particle is generally unlikely to exit to the right when the interaction is very strong.

The effect of possible quasi-bound states is again visible in the attractive, moderately strong interaction potential case of Fig. \ref{fig:mod_strength_n5}(c-f). Sharp peaks arise and the structure of $p_n$ is highly irregular.

The overall structure for $n_0=5$ is significantly more complex than that for $n_0=1$; the confined particle's energy may decrease or increase depending on $Lk_0$. Outcomes in which the confined particle loses energy arise for even very small $Lk_0$, whereas if the particle begins in the ground energy state with $n_0=1$, there is a range of small $Lk_0$ for which the particle remains in the ground state with probability 1.

\section{Convergence of numerical method}
\begin{figure}
  \centerline{\includegraphics[width=\textwidth]{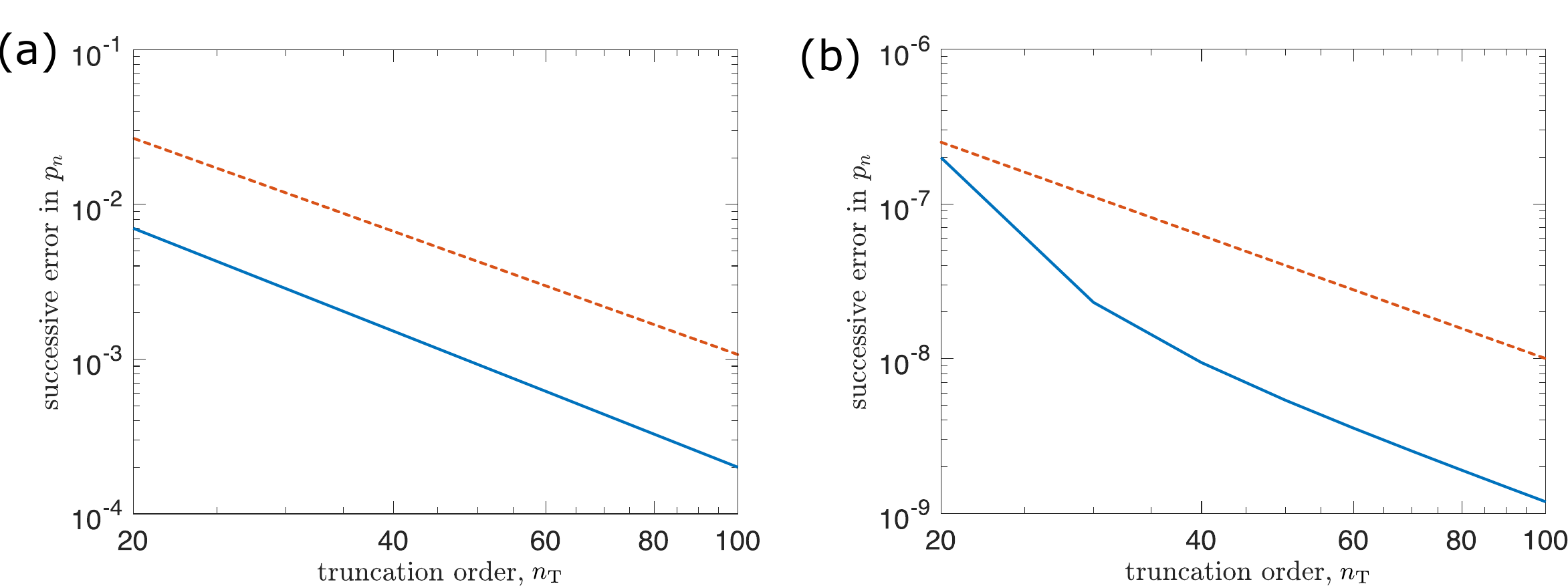}}
  \caption{\label{fig:convergence} Second-order pointwise convergence of numerical method. Orange dashed curves indicate lines with slope $-2$. (a) Arbitrarily chosen parameters $Lk_0=20$ and $\epsilon=4$ are fixed, and the number of mesh points $N$ varies with the truncation order $n_{\text{T}}$ by $N=2n_{\text{T}}$. Successive refinement in $n_{\text{T}}$ yields second-order convergence (slope $-2.08$). (b) Fixed $Lk_0=5.44141\gtrapprox \sqrt{3}\pi$ and $\epsilon=0.0045$ are fixed, to align with the parameters of Fig. \ref{fig:born_comparison}(c). Successive refinement in $n_{\text{T}}$ yields second-order convergence (terminal slope $-2.10$).}
\end{figure}

Here we demonstrate that the numerical method presented to solve (\ref{22}) exhibits second-order convergence with respect to simultaneous refinement in the truncation order $n_{\text{T}}$ and the number of mesh points $N$. We arbitrarily fix $Lk_0=20$ and $\epsilon=4$, thus assigning a moderate interaction strength, and compute the set of probabilities of all possible outcomes $p_n$ for $n_{\text{T}}=10,20,\dots 100$. We relate $N$ to $n_{\text{T}}$ via $N=2n_{\text{T}}$. We then plot the quantity 
\begin{equation}
\rvert p_n^{i+1}-p_n^i\rvert,\quad\quad i=1,\dots 9
\end{equation}
against $n_{\text{T}}$ on a log-log plot in Fig. \ref{fig:convergence}(a). The slope of the line is $-2.08$, which confirms second-order convergence of the numerical method. 


\section{Comparison with the Born approximation}

\begin{figure}
  \centerline{\includegraphics[width=\textwidth]{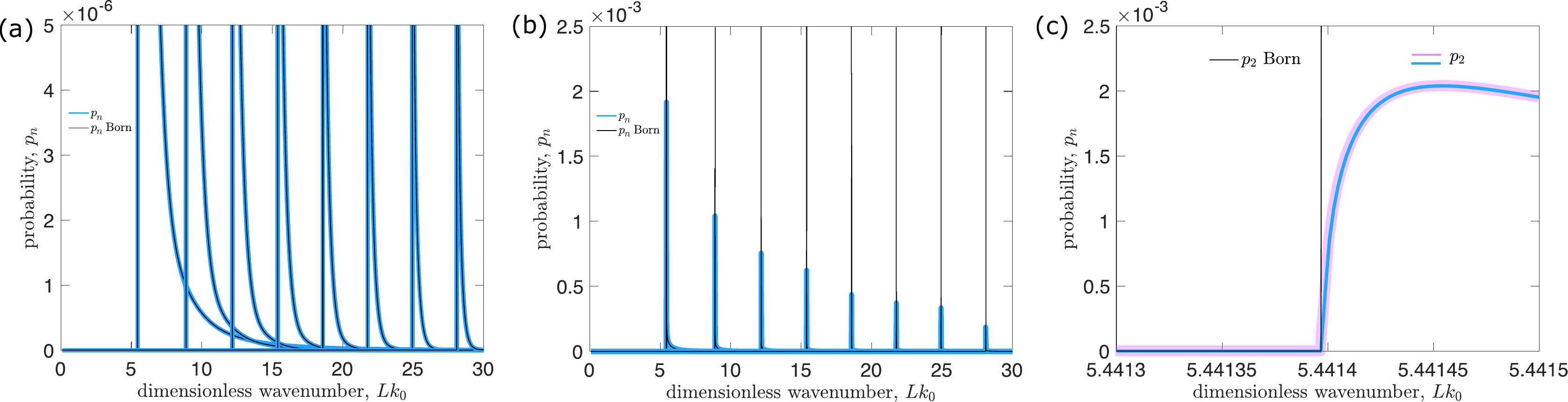}}
  \caption{\label{fig:born_comparison} Comparison of outcome probabilities computed via the Born approximation (black curves) and the full numerical solution (colored curves). Here, $n_0=1$ and the parameter $\epsilon$ ranges from $\mathcal{O}(10^{-1}) - \mathcal{O}(10^{-3})$. (a,b) Spectra of outcome probabilities exhibiting general agreement between the Born solution and the numerical solution. Panel (a) involves a truncated vertical axis to demonstrate agreement away from discontinuities. (c) Inset of panel (b), zoomed in the horizontal axis to the first peak for $p_2$. The Born approximation solution exhibits a true discontinuity and diverges to infinity at $k_0L=\sqrt{3}\pi\approx 5.4414$, whereas the numerical solution remains continuous and is convergent (as demonstrated by Fig. \ref{fig:convergence}(b)). The pink curve is associated with $n_T=10$, while the blue curve is associated with $n_T=50$. Both curves use $N = 2n_T$.}
\end{figure}

Solving (\ref{22}) using the Born approximation \cite{landau2013quantum} yields closed-form approximate solutions for outcome probabilities $p_n$ (see Appendix A). This perturbative approach aligns with the full computational solution in the weak-interaction limit as $\epsilon\to 0$. To illustrate this, Fig. \ref{fig:born_comparison} overlays solutions from the Born approximation atop the full solution for $n_0=1$ with a low interaction strength characterized by $\epsilon$ ranging from $\mathcal{O}(10^{-1}) - \mathcal{O}(10^{-3})$. Evidently by Fig. \ref{fig:born_comparison}(a), the two solutions match, and we note that in general, the Born approximation displays second-order convergence to the numerical solution as $\epsilon\to 0$. This agreement validates both approaches.

However, the high end of this interrogated range, $\epsilon\sim 10^{-1}$, is the maximal interaction strength such that the Born approximation remains valid. For larger $\epsilon$, the closed-form approximate solution cannot satisfy the requirement $\sum p_n = 1$. This restriction to weak interaction strengths is a fundamental limitation of such perturbative methods, which the scheme developed in this work overcomes.

Moreover, at any interaction strength $\epsilon$, no matter how small, the Born approximation invalidates itself by predicting transition probabilities greater than one and indeed approaching $+\infty$ over certain intervals of incident wave number of the free particle (Fig. \ref{fig:born_comparison}(b)).  As explained more fully in the Appendix, the problematic situation is one in which the incident particle has just slightly more energy than that needed to promote the confined particle from one energy level to a higher energy level.  If the transition were to occur, it would leave the free particle with very little energy, which would give it more time to interact with the confined particle, and this seems to strengthen the interaction to the point that the Born approximation yields results that do not make sense.  Our method does not have difficulty in this situation because it is not restricted to weak interaction (Fig. \ref{fig:born_comparison}(c)).  Fig. \ref{fig:born_comparison} shows both the excellent agreement our numerical solution holds with the Born approximation where that approximation is applicable, and also that our method can handle those special cases in which the Born approximation completely fails even though the interaction parameter $\epsilon$ is very small.



\section{Summary and conclusions}

In this paper, we have considered the one-dimensional scattering
of a free particle against a confined particle, in which the two
particles interact via a delta-function potential which may be
repulsive or attractive.  The confined particle is restricted
to move in the interval $(0,L)$, while the space available to the
free particle is the whole real line, and the boundaries $x=0$
and $x=L$ are transparent to the free particle.

Starting from the time-independent Schr{\"o}dinger equation, we
have derived an infinite system of integral equations, which
we have solved numerically by truncation and discretization.
The computational results show that the method is second-order
accurate.

We have considered an incoming product state in which the
free particle is incident from the left with prescribed energy,
and the confined particle is in one of the possible stationary
states of a particle in a one-dimensional box. We have
calculated the probabilities of the finitely many possible
outcomes, each of which corresponds to the confined particle
being in some (possibly different) stationary state of a particle
in a one-dimensional box, and the free particle flying out to
the right or left with some definite energy.  It is a direct
consequence of conservation of energy that there are only
finitely many possible outcomes.  In some of these outcomes,
the free particle loses energy and the particle in the box
moves up by one or more energy levels, but this is only possible
if the free particle has sufficient incoming energy.  In other outcomes, the free particle stimulates a loss of energy of
the confined particle, provided that the confined particle
was not in its ground state initially.

We have presented computational results which show how the probabilities of these discrete outcomes depend on the dimensionless wavenumber of the free particle for low, moderate, and high delta-potential interaction strengths, in both the attractive and repulsive cases. These representative examples cover two types of cases:  one
with the confined particle initially in its ground state, $n_0 = 1$, and another with the confined particle initially in an excited state, specifically $n_0 = 5$.

As demonstrated by our direct comparison against the Born approximation, the computational method developed here holds multiple distinct advantages against perturbative methods: first, it remains valid for arbitrary interaction strengths, while the Born approximation completely fails for sufficiently large interaction strengths; and second, even for sufficiently small interaction strengths such that the Born approximation is generally valid, our approach converges across the entire range of incident wave numbers of the free particle, while the Born approximation diverges over certain intervals.

An important motivation of this work has been to understand the nature
of state change in quantum mechanics. To this end, we have developed a
general framework which avoids the use of perturbation theory that can be used to calculate the probabilities of different asymptotic outcomes in one-dimensional scattering. Note that the type of scattering that we have considered involves the possibility of state change in the system that does the scattering, with that state change being induced by interaction the particle that is being scattered.

Whether the computational framework that we have developed can be of
 use, for example, in the study of quantum wires \citep{vaishnav2006hall, goncharov2014semispectral} and quantum
 dots \citep{jacak2013quantum, kouwenhoven1998quantum}, remains to be seen.  The idealizations discussed in
 the introduction may require modification to make such applications
 realistic, but we note in particular that the idealization of an
 infinitely deep potential well is one that is used, at least as a
 first approximation, in the formulation of simple models of quantum
 wires and quantum dots \citep{harrison2016quantum}.


\ack{We thank Michael I. Weinstein for helpful discussions.}

\funding{We acknowledge support from the Henry H. MacCracken Program at New York University.}


\data{All code used to generate data in this work is publicly available at \url{https://github.com/oliviapomerenk/State-Change-via-Scattering}.}

\section*{Appendix A}
Here we develop an approximate solution to (\ref{22}) in the weak-interaction limit using perturbation theory. In the Born approximation \cite{landau2013quantum}, equation (\ref{22}) becomes a formula for $\Phi_n(x)$:
\begin{equation}
    \Phi_n(x) = -\int_0^LG_n(x-x')A_{nn_0}(x')e^{ik_0x'}dx'.
\end{equation}
Note that this limiting form is reasonable only when the dimensionless quantity $\epsilon$ is small. In order to compute the probabilities $p_n^{\pm}$, it is only necessary to evaluate this at $x=0$ and $x=L$, where it becomes
\begin{align}
    \Phi_n(0) &= -\int_0^LG_n(-x)A_{nn_0}(x)e^{ik_0x}dx, \\
    \Phi_n(L) &= -\int_0^LG_n(L-x)A_{nn_0}(x)e^{ik_0x}dx'.
\end{align}
Since it must be that $c_n<0$, $G_n(x)$ is given by (\ref{Gn_k}). Substituting this as well as the definition for $A_{nn_0}$ given by (\ref{Ann}) yields
\begin{align}
    \Phi_n(0) &= \frac{1}{2ik_n}\frac{4m_1\mu_0}{\hbar^2 L}\int_0^L\sin\left(\frac{n\pi x}{L}\right)\sin\left(\frac{n_0\pi x}{L}\right)e^{i(k_0+k_n)x}dx, \label{60} \\
    \Phi_n(L) &= \frac{e^{ik_nL}}{2ik_n}\frac{4m_1\mu_0}{\hbar^2 L}\int_0^L\sin\left(\frac{n\pi x}{L}\right)\sin\left(\frac{n_0\pi x}{L}\right)e^{i(k_0-k_n)x}dx \label{61}
\end{align}
where $k_n$ is given by (\ref{kn_def}) with $c_n$ given by (\ref{an_def}). From here on we assume
 that $m_1 = m_2$, and in that case the formula for $k_n$ is
 \begin{equation}
     k_n = \sqrt{k_0^2 - \left(\frac{\pi}{L}\right)^2\left(n^2-n_0^2\right)}.
 \end{equation}
 If $n > n_0$, the argument of the square root in the above can be negative. This will happen if the free particle does not have enough energy to bring about the transition from state $n_0$ to the higher-energy state $n$ of the confined particle.  In that case, the transition cannot occur. Therefore, the formulae that follow are only meaningful when $k_n$ is
 real, and the transition probabilities stated below need to be
 understood to be zero when $k_n$ is imaginary.  There is an additional difficulty when $k_n$ is real but small, and this will be discussed below.

Continuing now with the case $m_1=m_2$, we can evaluate the integrals in (\ref{60}-\ref{61}) by making use of trigonometric identities to obtain
\begin{align}
    \Phi_n(0) &= \begin{dcases}
        -\epsilon\frac{k_0}{k_n}\frac{1-e^{i(k_0+k_n)L}}{(k_0+k_n)L}\frac{n_0n}{\left(\frac{k_0L}{\pi}\right)^2-n^2},\quad n_0\pm n\text{ even} \\
        -\epsilon\frac{k_0}{k_n}\frac{1+e^{i(k_0+k_n)L}}{(k_0+k_n)L}\frac{n_0n}{\left(\frac{k_0L}{\pi}\right)^2-n^2},\quad n_0\pm n\text{ odd.}
    \end{dcases} \\
      \Phi_n(L) &= \begin{dcases}
        \epsilon\frac{k_0}{k_n}\frac{e^{ik_nL}-e^{ik_0L}}{(k_0-k_n)L}\frac{n_0n}{\left(\frac{k_0L}{\pi}\right)^2-n^2},\quad n_0\pm n\text{ even, }n\neq n_0 \\
        \epsilon\frac{k_0}{k_n}\frac{e^{ik_nL}+e^{ik_0L}}{(k_0-k_n)L}\frac{n_0n}{\left(\frac{k_0L}{\pi}\right)^2-n^2},\quad n_0\pm n\text{ odd}
    \end{dcases}  
\end{align}
where $\epsilon=m\mu_0/k_0\hbar^2$ is the dimensionless quantity which encodes the strength of the particle-particle interaction relative to the dimensionless wavenumber of the incident particle (equation (\ref{epsilon})). Note the restriction for $\Phi_n(L)$ in the case $n_0\pm n$ even, for which $n$ must not equal $n_0$. This is because, due to the condition $m_1=m_2$, if $n=n_0$ then $k_n=k_0$ (this does not present a division by zero issue for the case $n_0\pm n$ odd, since then it cannot be true that $n=n_0$). 

Using definitions (\ref{pnplus}) and (\ref{pnminus}) and further simplifying, we may write using the above
\begin{align}
    p_n^- &= \begin{dcases}\label{pnminusborn}
        \epsilon^2\frac{k_0}{k_n}\left(\frac{\sin\left(\frac{(k_0+k_n)L}{2}\right)n_0n}{\frac{(k_0+k_n)L}{2}\left[\left(\frac{k_0L}{\pi}\right)^2-n^2\right]}\right)^2 & n_0 \pm n \text{ even} \\
        \epsilon^2\frac{k_0}{k_n}\left(\frac{\cos\left(\frac{(k_0+k_n)L}{2}\right)n_0n}{\frac{(k_0+k_n)L}{2}\left[\left(\frac{k_0L}{\pi}\right)^2-n^2\right]}\right)^2 & n_0 \pm n \text{ odd.}
    \end{dcases} \\
    p_n^+ &= \begin{dcases}\label{pnplusborn}
        \epsilon^2\frac{k_0}{k_n}\left(\frac{\sin\left(\frac{(k_0-k_n)L}{2}\right)n_0n}{\frac{(k_0-k_n)L}{2}\left[\left(\frac{k_0L}{\pi}\right)^2-n^2\right]}\right)^2 =  & n_0 \pm n \text{ even, } n_0\neq n \\
        \epsilon^2\frac{k_0}{k_n}\left(\frac{\cos\left(\frac{(k_0-k_n)L}{2}\right)n_0n}{\frac{(k_0-k_n)L}{2}\left[\left(\frac{k_0L}{\pi}\right)^2-n^2\right]}\right)^2 & n_0 \pm n \text{ odd.}
    \end{dcases}
\end{align}
As mentioned above, these formulae are to be interpreted as zero when $k_n$ is imaginary, and there is an additional difficulty when $k_n$ is real but small.  This additional difficulty is due to the factor $k_0/k_n$ that multiplies $\epsilon^2$ in all of the formulae for the transition probabilities.  There is nothing to prevent $k_0/k_n$ from being arbitrarily large, and when this happens the interaction of the two particles is no longer weak, and the Born approximation becomes invalid.  The physical situation in which this occurs is one in which the free particle has just slightly more energy than the amount needed to cause a particular upward change of state of the confined particle.  If such a transition does occur, it leaves the free particle with almost no energy, so it stays in the vicinity of the confined particle for a long time, thus increasing the effective strength of the interaction.  Indeed, the Born approximation invalidates itself in this situation, since it predicts transition probabilities that are greater than one, and moreover those probabilities approach $+\infty$ as $k_n\to 0$.
 
To compute $p_n^+$ when $n_0=n$, we simply use the fact that these finitely many probabilities must sum to 1:
\begin{equation}
    p_{n_0}^+ = 1 - \left(p_{n_0}^- +\sum_{n:c_n<0, n\neq n_0}(p_n^+ + p_n^-)\right)
\end{equation}

In (\ref{pnminusborn}) and (\ref{pnplusborn}), the case $k_0L/\pi=n$ appears problematic. But, it may be shown that $k_0L/\pi=n\Rightarrow k_nL/\pi=n_0$, and so $(k_0\pm k_n)L/2 = \pm (n_0-n)\pi/2$. It follows that when $k_0L/\pi=n$, $\sin\left(\frac{(k_0-k_n)L}{2}\right)=0$ when $n_0\pm n$ is even and $\cos\left(\frac{(k_0-k_n)L}{2}\right)=0$ when $n_0\pm n$ is odd. So, this special value of $k_0$ gives a finite result which may be evaluated by L'H\^{o}pital's rule.

\bibliography{iopjournal-guidelines}

\end{document}